\pdfoutput=1

\documentclass[twocolumn, times]{aastex61}
\usepackage{natbib}
\usepackage{array,multirow}

\newcommand{\nii}{[$\mbox{N\,II}$]\,}
\newcommand{\cii}{[$\mbox{C\,II}$]\,}
\newcommand{\oii}{[$\mbox{O\,II}$]\,}
\newcommand{\sii}{[$\mbox{S\,II}$]\,}
\newcommand{\um}{\,$\mu$m\,}

\received{}
\revised{}
\accepted{}

\submitjournal{ApJL}

\shorttitle{First [NII] 122\um detection at $z > 4$}
\shortauthors{Minju M. Lee et al.}
\defcitealias{Lu2017b}{Lu17}

\begin{document}

\title{First [NII]122 \um line detection in a QSO-SMG pair BRI 1202-0725 at z=4.69}

\correspondingauthor{Minju M. Lee}
\email{minju@mpe.mpg.de}

\author[0000-0002-2419-3068]{Minju M. Lee}
\affiliation{Max-Planck-Institut f\"{u}r Extraterrestrische Physik (MPE), Giessenbachstr., D-85748 Garching, Germnay}
\affiliation{Division of Particle and Astrophysical Science, Graduate School of Science, Nagoya University, Furo-cho, Chikusa-ku, Nagoya 464-8602, Japan}
\affiliation{National Observatory of Japan, 2-21-1 Osawa, Mitaka, Tokyo 181-0015, Japan}

\author{Tohru Nagao}
\affiliation{Graduate School of Science and Engineering, Ehime University, 2-5 Bunkyo-cho, Matsuyama 790-8577, Japan}
\author{Carlos De Breuck}
\affiliation{European Southern Observatory, Karl Schwarzschild Stra{\ss}e 2, 85748 Garching, Germany}

\author{Stefano Carniani}
\affiliation{Scuola Normale Superiore, Piazza dei Cavalieri 7, I-56126 Pisa, Italy}
\author{Giovanni Cresci}
\affiliation{INAF - Arcetri Observatory, Florence, Italy}
\author{Bunyo Hatsukade}
\affiliation{Institute of Astronomy, Graduate School of Science, The University of Tokyo, 2-21-1 Osawa, Mitaka,
Tokyo 181-0015, Japan}
\author{Ryohei Kawabe}
\affiliation{National Astronomical Observatory of Japan, 2-21-1 Osawa, Mitaka, Tokyo 181-8588, Japan}
\affiliation{SOKENDAI (The Graduate University for Advanced Studies), 2-21-1 Osawa, Mitaka, Tokyo 181-8588,
Japan}
\affiliation{Department of Astronomy, The University of Tokyo, 7-3-1 Hongo, Bunkyo-ku, Tokyo 113-0033, Japan}
\author{Kotaro Kohno}
\affiliation{Institute of Astronomy, Graduate School of Science, The University of Tokyo, 2-21-1 Osawa, Mitaka,
Tokyo 181-0015, Japan}
\affiliation{Research Center for the Early Universe, The University of Tokyo, 7-3-1 Hongo, Bunkyo, Tokyo 113-0033, Japan}

\author{Roberto Maiolino}
\affiliation{Cavendish Laboratory, University of Cambridge, 19 J. J. Thomson Avenue, Cambridge CB3 0HE, UK}
\affiliation{Kavli Institute for Cosmology, University of Cambridge, Madingley Road, Cambridge CB3
0HA, UK}
\author{Filippo Mannucci}
\affiliation{INAF—Osservatorio Astrofisico di Arcetri, Largo E. Fermi 5, 20125 Firenze, Italy}
\author{Alessandro Marconi}
\affiliation{INAF—Osservatorio Astrofisico di Arcetri, Largo E. Fermi 5, 20125 Firenze, Italy}
\affiliation{Dipartimento di Fisica e Astronomia, Universit\'{a} degli Studi di Firenze, Via G. Sansone 1, 50019 Sesto F.no, Firenze, Italy}
\author{Kouichiro Nakanishi}
\affiliation{National Observatory of Japan, 2-21-1 Osawa, Mitaka, Tokyo 181-0015, Japan}
\affiliation{SOKENDAI (The Graduate University for Advanced Studies), 2-21-1 Osawa, Mitaka, Tokyo 181-8588,
Japan}
\author{Toshiki Saito}
\affiliation{Max-Planck Institute for Astronomy, K\"{o}nigstuhl, 17 D-69117 Heidelberg, Germany}
\author{Yoichi Tamura}
\affiliation{Division of Particle and Astrophysical Science, Graduate School of Science, Nagoya University, Furo-cho, Chikusa-ku, Nagoya 464-8602, Japan}

\author{Paulina Troncoso}
\affiliation{Universidad Universidad Aut\'{o}noma de Chile, Chile. Av. Pedro de Valdivia 425, Santiago, Chile}

\author{Hideki Umehata}
\affiliation{RIKEN Cluster for Pioneering Research, 2-1 Hirosawa, Wako, Saitama 351-0198, Japan}
\author{Min Yun}
\affiliation{Department of Astronomy, University of Massachusetts, Amherst, MA 01003, USA}



\begin{abstract}
We report the first detection obtained with ALMA of the \nii 122\um line emission from a galaxy group BRI 1202-0725 at $z=4.69$ consisting of a QSO and a submilimeter-bright galaxy (SMG). 
Combining with a detection of \nii 205\um line in both galaxies, we constrain the electron densities of the ionized gas based on the line ratio of \nii122/205.
The derived electron densities are $26^{+12}_{-11}$ and $134^{+50}_{-39}$ cm$^{-3}$ for the SMG and the QSO, respectively.
The electron density of the SMG is similar to that of the Galactic Plane and to the average of the local spirals. 
Higher electron densities by up to a factor of three could however be possible for systematic uncertainties of the line flux estimates. 
The electron density of the QSO is comparable to high-$z$ star-forming galaxies at $z=1.5-2.3$, obtained using rest-frame optical lines and with the lower limits suggested from stacking analysis on lensed starbursts at $z=1-3.6$ using the same tracer of \nii.
Our results suggest a large scatter of electron densities in global scale at fixed star formation rates for extreme starbursts.
The success of the \nii 122\um and 205\um detections at $z=4.69$ demonstrates the power of future systematic surveys of extreme starbursts at $z>4$ for probing the ISM conditions and the effects on surrounding environments.
\end{abstract}




\section{Introduction}\label{sec:intro}
Understanding the physical conditions of star formation is critical in constraining theoretical models of galaxy evolution. 
Galaxies form stars at a higher rate in the early universe at a fixed mass.
A following question is how the interstellar medium (ISM) properties are correspondingly changed to understand the cosmic evolution.
Observations of $z>1$ star-forming galaxies suggest that the ISM state and/or the hardness of the extreme ultraviolet (EUV) radiation field were more extreme in the past than in the present day.
For example, rest-frame optical line observations revealed that electron densities of high redshift star-forming galaxies range between $100-1000$ cm$^{-3}$, which is up to two orders of magnitude higher than those observed in the local universe (e.g., \citealt{Masters2014, Steidel2014, Sanders2016, Kaasinen2017}).

Far-infrared transitions are a powerful tool for investigating the ISM. 
The fact that they are less affected by dust 
compared to optical line tracers is a strong advantage to use them.
 At wavelengths greater than 100~$\mu$m, the fine-structure transitions of \cii157.7\um, the \nii121.9\um and 205.2\um have been used for probing ISM conditions of local and high-$z$ galaxies (\citealt{Wright1991, Stacey1991, Lord1996, Bennett1994, Malhotra2001, Brauher2008, Nagao2012, Farrah2013, Zhao2013, Zhao2016a, Zhao2016b, Herrera-Camus2016, Herrera-Camus2018a, Herrera-Camus2018b}). 
With an ionization threshold of 11.3 eV, the \cii line emission arises from the neutral and the ionized gas. 
On the other hand, the two [NII] fine-structure lines originate from fully ionized gas since the ionization potential of nitrogen (14.5 eV) is about $\sim 0.9$ eV higher than that of hydrogen.
 Therefore, the ionized nitrogen \nii lines reflect the effect of UV photons emitted by massive young stars, with possible enhancement from X-ray photoionization.
 The combination of two fine structure lines can be used as a tracer of electron density and this diagnostic barely depends on the electron temperature (e.g., \citealt{Goldsmith2015, Herrera-Camus2016}).

The \nii122\,\um line emission has not been detected for galaxies at $4<z<7$ till now, which is the epoch when larger number of galaxies are beginning to form after the end of the reionization. 
In this Letter, we report the first detection of \nii122\,\um line from a QSO-SMG pair, BRI 1202-0725, at $z=4.69$.
This compact group of BRI 1202−0725 was one of the first $z > 4$ submillimeter-bright systems discovered (\citealt{Isaak1994}) and remains the archetype for major starbursts in gas-rich mergers in the early universe. 
It consists of an optically selected QSO, optically faint SMG, which is located $4'' (\approx$26 kpc) northwest of the quasar (\citealt{Omont1996, Hu1996}), and two Lyman-$\alpha$-selected galaxies in their very vicinity (\citealt{Hu1996, Fontana1996, Ohta2000, Salome2012, Carilli2013b, Carniani2013})\edit1{\deleted{ -- one between QSO and SMG and the other at $\sim2''$ northwest of the QSO}}.
Extremely high FIR luminosities of QSO and SMG (\citealt{Omont1996, Iono2006, Yun2000}) ($\sim10^{13}~L_{\odot}$) imply vigorous star forming activity of $\approx1000$ $M_{\odot}$ yr$^{-1}$.
The system is known to have rich C-bearing emission line data sets;
various rotational CO molecular lines have been detected to up $J = 11$ (e.g., \citealt{Ohta1996, Omont1996, Salome2012}) in addition to bright \cii emissions (\citealt{Iono2006, Carilli2013b}).
\cite{Lu2017b} (hereafter, \citetalias{Lu2017b}) reported the first detection of \nii205\,$\mu$m line emissions for both systems and measured the dust temperature ($T_{\rm dust}=43{\pm}2$ K) using the line ratio between the [NII] line and CO~(7--6).
We add new \nii122\,\um line detections, which provide further constraints on the physical conditions of the ISM, namely the electron density.

\edit1{\deleted{This paper is laid out in the following manner. In Section~\ref{sec:obs}, we describe the ALMA observations and data analysis process. In Section~\ref{sec:results}, we present the detection of the \nii lines. In Section~\ref{sec:discussion}, we discuss electron densities estimated from \nii 122/205 line ratio.
We present our conclusion in in Section~\ref{sec:conclusion}.}}
We assume $H_0 = 67.8$ ${\rm km\,s^{-1}\,Mpc^{-1}}$, $\Omega_0 = 0.308$ and $\Omega_{\Lambda} = 0.692$ \citep{PlanckCollaboration2015}.

\section{ALMA Observations and data reduction}\label{sec:obs}
\subsection{Band 6 observations for [NII]~205~$\mu$m}
The Band 6 observations were carried out for our ALMA Cycle 2 program.
A total of 39 and 40 antennas were used with the unprojected length ($L_{\rm baseline}$) between 15--348 m (C34-2/1)
on 2014 December 14 and 2015 January 4 with the total on-source time of 58 minutes.

We used four spectral windows (SPW), each of 1.875 GHz wide. Two of them were set in the upper sideband with 3.906 MHz resolution ($\sim$ 4.5 km s$^{-1}$) to target [NII]~205~$\mu$m. One SPW in the lower side band was also set to 3.906 MHz resolution. The remaining SPW was set to 7.812 MHz resolution ($\sim$ 9.7 km s$^{-1}$), in which we detected CO~(12--11) emissions both from the SMG and the QSO (Lee et al. 2019, in preparation).
Two quasars, J1256-0547 and J1058+0133, were chosen for the bandpass calibration. 
J1216-1033 was the phase calibrator.
Callisto was the flux calibrator.

\subsection{Band 8 observations for [NII]~122~$\mu$m}
The Band 8 observations at 700~$\mu$m were also a subset of the same ALMA Cycle 2 program.
Observations used 37 or 38 antennas with the unprojected length ($L_{\rm baseline}$) between 21--783 m on 2015 June 6 through 8 and total on-source time was 112 minutes.

We used four spectral windows (SPW), each of 1.875 GHz wide. Two of them were set in the upper sideband with 3.906 MHz resolution ($\sim$ 2.7 km s$^{-1}$) to detect \nii122\um. The spectral resolution for the remaining two SPWs in the lower sideband was set to 7.812 MHz ($\sim$ 5.6 km s$^{-1}$).
J1256-0547 was chosen as a bandpass and a phase calibrator.
3C~273 and Titan were chosen for flux calibration.

\subsection{Archival data : Band 6 archival data}
We downloaded the archival data sets that were independently taken during ALMA Cycle 3 for \nii 205\um line detection reported in \citetalias{Lu2017b}. 
The details of the observations are presented in \citetalias{Lu2017b}.
We calibrated the data based on the provided pipeline script.
It was observed in the time-domain mode (TDM) with a spectral resolution of 15.625 MHz, corresponding to $\sim$ 19 km s$^{-1}$, in which the spectral sampling is a factor of $\approx4$ coarser than our Band 6 data sets. 
Hereafter, we name the data as the ``Lu data".

\subsection{Data reduction and analysis}
We performed calibration using the Common Astronomy Software Applications package ($\mathtt{CASA}$, 
\citealt{McMullin2007}).
For our Band 6 and 8 data sets, we used the calibration scripts provided by the ALMA ARC members that used CASA versions of 4.2.2 and 4.3.1, respectively. For the Lu data, we used the CASA version 4.5.2.

Images were produced by $\mathtt{CASA}$ task $\mathtt{tclean}$.
All imaging processes were handled with version 5.4.0.
Using the natural weighting, the synthesized beam sizes are $1''.43\times0''.84$ and $0''.32\times0''.24$ for the [NII] 205\,$\mu$m and [NII]~122$\mu$m observations, respectively.
For the Lu data, the beam size is $0''.97\times0''.80$.

Provided the different resolutions obtained in different bands, we tried to match the resolutions as much as possible.
To compare with the Lu data, we made $1''.5$-resolution images for all Band 6 data sets and estimated the line widths and fluxes. 
For the \nii122\um data, we investigated the S/Ns over a few $\mathtt{uvtaper}$ parameters.
We chose the $uv$-tapering parameter of 330k$\lambda$ with the synthesized beam of $0''.44\times0''.38$, which is the size without losing significant S/N, i.e., peak S/N from $\approx 7.3 (7.3)$ to $\approx$7.1 (8.2) for the SMG(QSO).
We also made heavily-tapered \nii122\um images to obtain a resolution close to the \nii205\um data. 
With the $uv$-tapering parameter of 80k$\lambda$, the beam size is $1''.20\times1''.13$. 
This gives lower peak S/Ns of $\sim$2-3 for both galaxies. 
In the discussion section, we use the highly tapered images (``80k$\lambda$-tapered map'') to evaluate potential systematic errors.
For our \nii~205\um data, we applied the Briggs weighting with a robustness parameter of 0.5, which gives a synthesized beam of $1''.32\times0''.68$.
We subtracted the continuum based on image data cube using $\mathtt{imcontsub}$ to control better the continuum shape especially for the targets away from the phase center and hence to get higher S/N than using $\mathtt{uvcontsub}$.
We checked that the flux measured from the data after applying $\mathtt{uvcontsub}$ gives consistent values within errors. 

We measure the flux after investigating the flux growth curves using various aperture sizes. 
The flux values reach the asymptotic values with aperture sizes of $1''.2$ and $3''.0$ for the \nii122\um and \nii205\um, respectively. Using these aperture sizes, we derived the flux values based on a Gaussian fit using the CASA task $\mathtt{imfit}$.

\subsection{Missing flux}
\begin{figure}
\center
\includegraphics[width = 7 cm, bb = 0 0 1000 800]{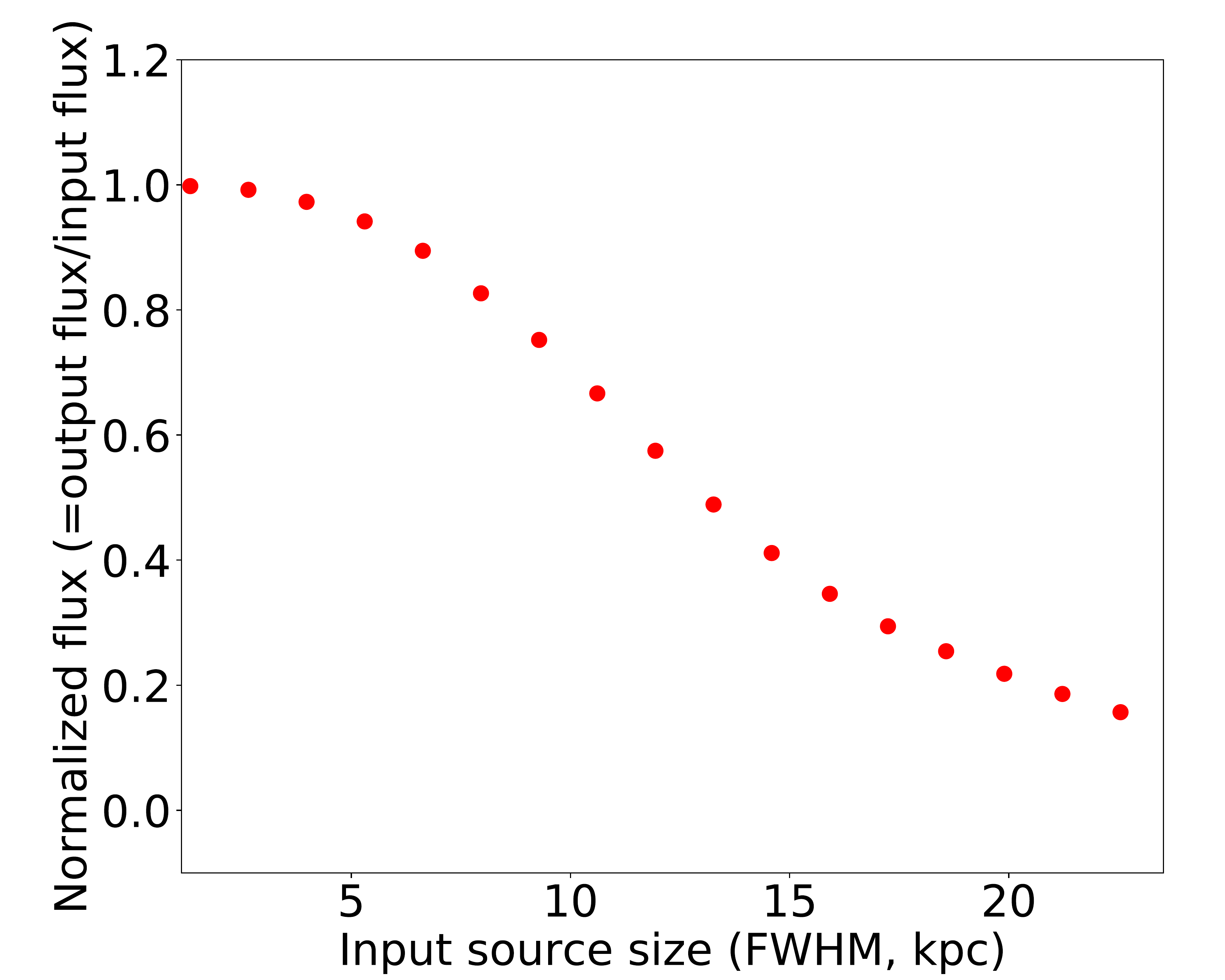}
\caption{The expected missing flux assuming a Gaussian distribution at given flux using CASA simulation with $uv$ tapering of 330k$\lambda$ for an ideal case without noise. \label{fig:missingflux}}
\end{figure}

Considering the high angular resolution obtained in the [NII] 122$\mu$m observations, we explored the possibility of emission from the extended regions.
We investigated the mock observations of a Gaussian structure component with various sizes at a given configuration of C34-5 during Cycle 2 using the CASA task $\mathtt{simobserve}$.
Figure~\ref{fig:missingflux} shows such an experiment when the images are created after applying the same $uv$-tapering parameter of 330k$\lambda$.
At an ideal condition of infinity S/N (i.e., without noise), we were able to recover the flux for more than 80\% of the input value, when the source is extended up to $\approx8$ kpc.

We estimated the sizes of [NII]-emitting regions using the CASA task $\mathtt{imfit}$. 
We used the natural weight maps of the \nii122\um. 
We could constrain the size only for the QSO, which is $0''.43(\pm0''.15) \times0''.23 (\pm0''.18)$ and obtain the upper limit for the SMG, which is $0''.38 \times0''.23$.
For comparison, the beam-deconvolved \nii205\um sizes are $0''.59 (\pm 0''.19) \times 0''.42(\pm 0''.21)$ and $0''.78 (\pm 0''.18) \times 0''.62(\pm 0''.38)$ for the QSO and the SMG, respectively from the briggs-weighted maps.
While there is a hint of smaller sizes for the \nii122 emissions compared to the \nii205 from these measurements, we note that the uncertainties are also large. 
At least from the Gaussian fit, we conclude that both \nii lines are emitted from the regions of similar sizes comparable to or smaller than the \cii-emitting regions, which are $\approx 2-3$ kpc in scale radius (\citealt{Carniani2013}). 
\citetalias{Lu2017b} reported extended emissions (i.e., $\sim$ 9 kpc ($\approx1''.4$) for the QSO and 14 kpc ($\approx2''.1$) for the SMG) in \nii205\um line, which are larger than the estimates from our data.
We could only constrain the \nii205\um size for the QSO from the Lu data, $0''.81(\pm0''.21)\times0''.49(\pm0''.34)$ which is consistent with our data (the size before deconvolution is $1''.21 (\pm0''.10)\times1.00 (\pm0''.07))$. 
For the SMG, the size before beam deconvolution is $1''.58 (\pm0''.19)\times0''.80 (\pm0''.06)$, but the fit gives only an upper limit of the size to be $1''.50\times 0''.28$.
It may be worth noting that our data is 1.4$\times$ deeper in terms of the point source sensitivity. 
Considering this, it is less likely that a significant amount of emission is coming from the extended regions ($>10 $kpc). 
Therefore, we rely on the flux measurements without any correction.

\section{Results}\label{sec:results}
\begin{figure*}
\center
\includegraphics[width = 16 cm, bb = 0 0 1024 1024]{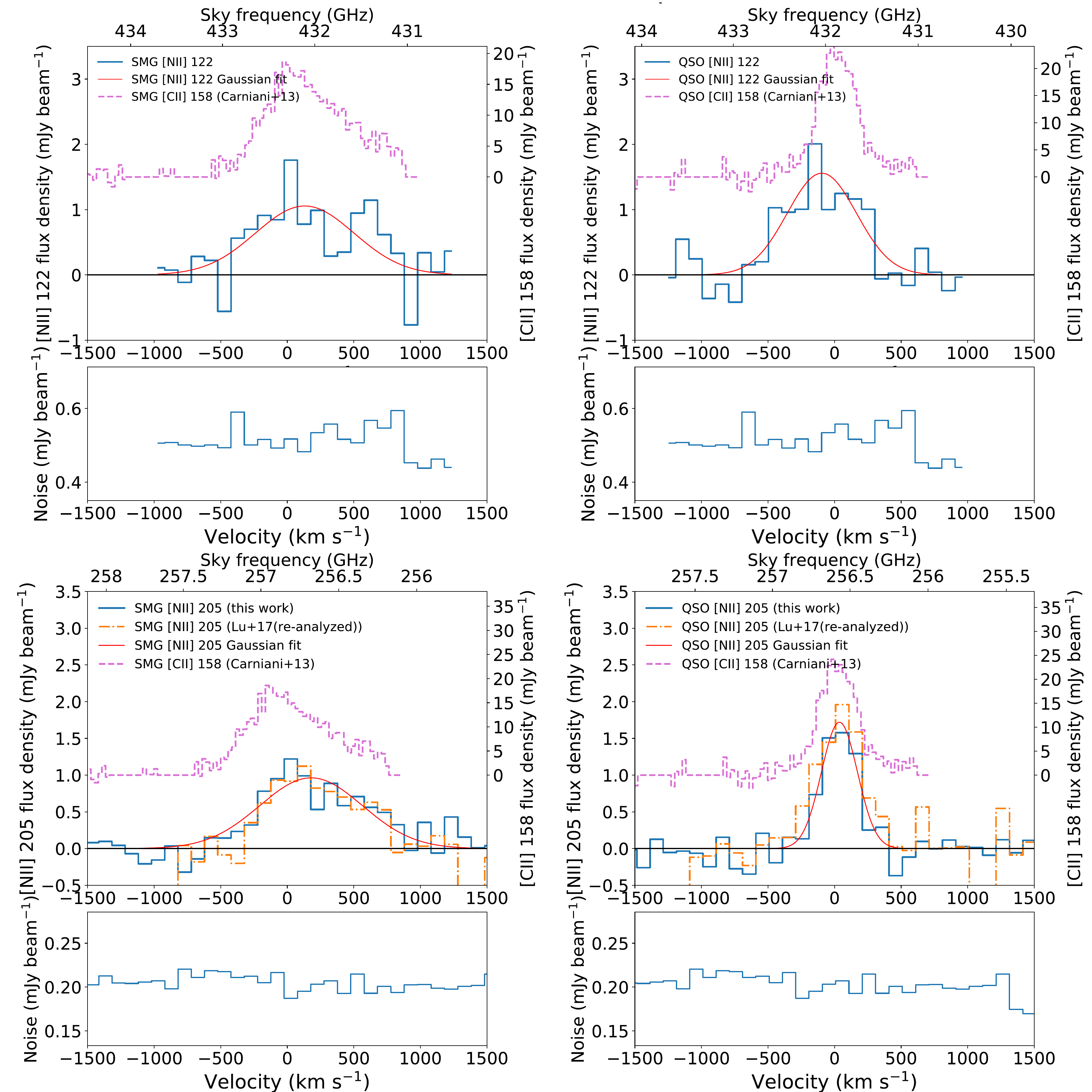}
\caption{Detection of \nii122 (top) and \nii205 (bottom) lines from the SMG (left) and the QSO (right) in blue solid lines. Top: The \nii122 spectra. The \nii122\,\um spectrum is extracted from the peak position using the the 330k$\lambda$-tapered cube  (i.e., $0''.44\times0''.38$) with 100 km s$^{-1}$ resolution.
We overplot \cii158\um line from \cite{Carniani2013} with the base level shifted to 1.5 for clarity.
Bottom: The \nii205 spectra in 100 km s$^{-1}$ resolution. The spectra are obtained from the peak positions using the $1''.5\times1''.5$ resolution cubes.
Overlaid orange dashed lines show the \nii205\um detection from another independent data set reported in \citetalias{Lu2017b} which we reanalyzed for comparison. We matched the resolution to have a synthesized beam of $1''.5\times1''.5$ for the Lu data as well. We overplot \cii158\um line from \cite{Carniani2013} with the base level shifted to 1.0 for clarity.
The velocity centers of the spectra are based on the redshifts from \cii\,158\um observations in \cite{Carilli2013b}. \label{fig:niiall}}
\end{figure*}

For the \nii205 line emission, we found our measurements are consistent with the Lu data within the uncertainties, in terms of the peak positions, line widths, and line luminosities. 
The spectra of all these data sets are shown in Figure~\ref{fig:niiall}.
\citet{Pavesi2016} used our Band 6 data and reported the flux measurement briefly, which we reconfirm the values using the same data set but with different analysis.
We note that \citetalias{Lu2017b} reported different flux values i.e., $0.99\pm0.02$ and $1.01\pm0.02$ for the SMG and the QSO, respectively, in which they used different aperture sizes for individual galaxies as opposed to ours. 
The flux values with the same flux extraction methods to ours using the Lu data are $1.07\pm0.16$ (SMG) and $0.81\pm0.10$ (QSO). 
These are consistent with our measurements listed in Table~\ref{tab:flux} within the errors. 
However, all flux values using the TDM data tend to be smaller(larger) for the SMG(QSO) compared to our data (in frequency-domain mode).
While it is difficult to investigate the origin of the difference, we emphasize that we measured the line fluxes after a careful analysis of flux growth curves and aperture photometry.

\begin{figure*}
\center
\includegraphics[width = 14 cm, bb = 0 0 1920  1600]{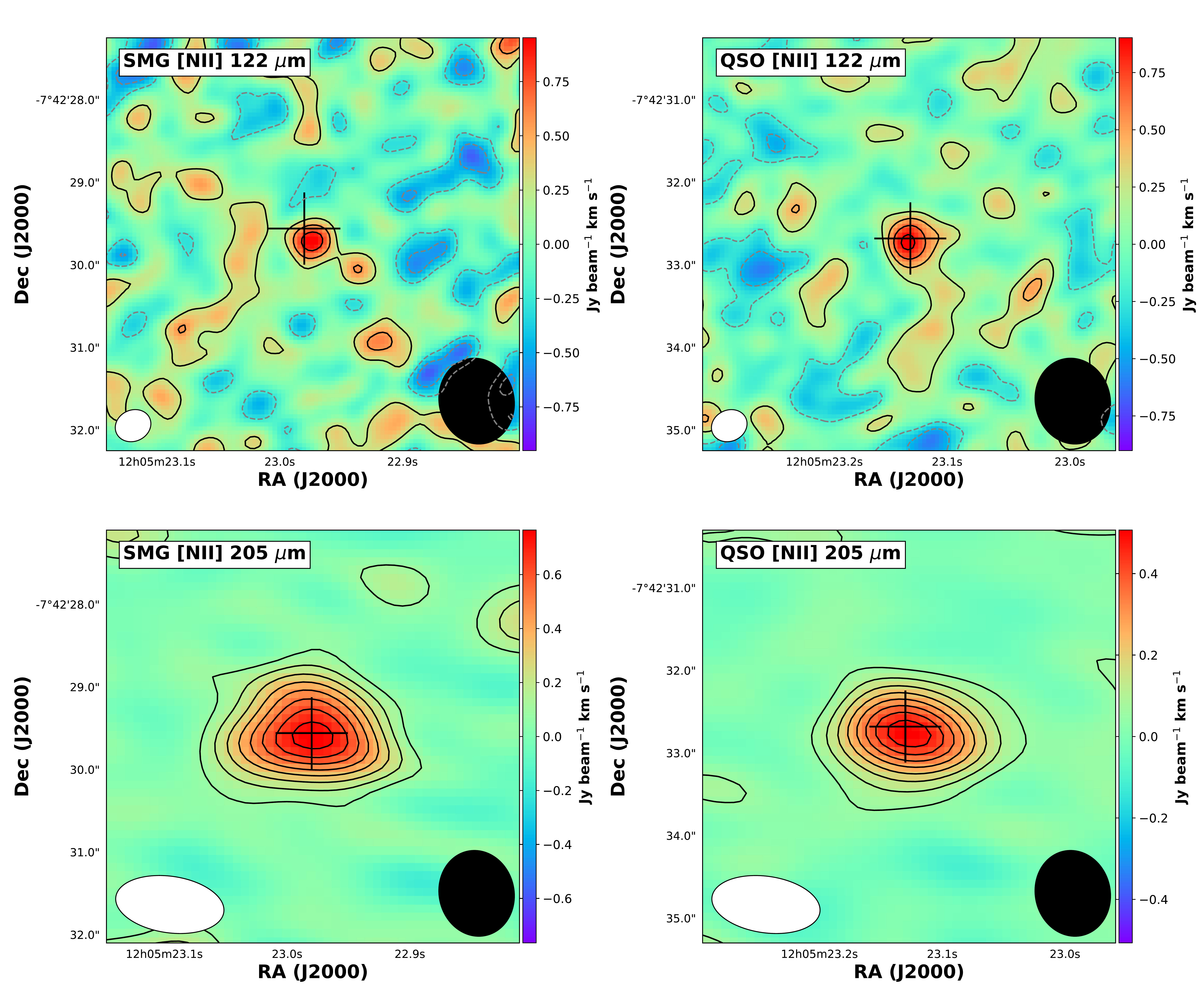}
\caption{Top: The line intensity of \nii122\,\um of the SMG (left) and the QSO (right). Contour lines are starting from 2$\sigma$, in steps of 2$\sigma$ where $1\sigma$ is 0.13 and 0.10 in Jy km s $^{-1}$ for the SMG and the QSO, respectively. We also added negative contours of -4$\sigma$ and -2$\sigma$ by gray dashed lines. The beam sizes after $uv$-tapering are shown by white filled ellipses, which is $0''.44\times0''.38$. Bottom: The line intensity of \nii205\,\um for the SMG (left) and the QSO (right), respectvely. Contour lines are starting from 2$\sigma$, in steps of 2$\sigma$ where $1\sigma$ is 0.04 and 0.03 in Jy km s $^{-1}$ for the SMG and the QSO, respectively. The beam size is $1''.32\times0''.68$. All panel sizes are $5''$ width. The cross markers are the peak positions of the \cii158 line and the ellipse filled in black is the beam size of the \cii observations, which is $0''.8\times0''.7$. \label{fig:nii122}}
\end{figure*}

\movetabledown=1.6in
\begin{deluxetable*}{ccccccccccccccc}
\tabletypesize{\scriptsize}
\tablecaption{Flux measurements\label{tab:flux}}
\tablewidth{0pt}
\tablehead{
\colhead{Target} &&\colhead{SMG} &&&&  \colhead{QSO} &\\
\cline{2-4}\cline{6-8}
	& \colhead{$F_{\rm line}\tablenotemark{a}$} & \colhead{FWHM} & \colhead{$L_{\rm line}$} & 
	&\colhead{$F_{\rm line}\tablenotemark{a}$} & \colhead{FWHM} & \colhead{$L_{\rm line}$} & \\
	&[Jy km s$^{-1}$]	& [km s$^{-1}$] & [$\times 10^9 L_{\odot}$]&
	& [Jy km s$^{-1}$]	& [km s$^{-1}$] & [$\times 10^9 L_{\odot}$]&\\
(1) & \colhead{(2)} & \colhead{(3)}	 & \colhead{(4)}	 & 
& \colhead{(5)} &\colhead{(6)} & \colhead{(7)}
}
\startdata
$[$NII$]$~122~$\mu$m	& $ 1.13 \pm 0.27$	&	$ 871 \pm 228$	&	$ 2.71\pm 0.65$&
					& $1.62 \pm 0.27$	&	$ 613\pm 133$	&	$ 3.89 \pm 0.65$	
					\\
$[$NII$]$~205~$\mu$m 	& $1.32 \pm 0.11$ & $1009\pm147$ & $1.86\pm0.15$ & 
								& $0.70\pm0.05 $ & $297\pm104$ & $0.98\pm 0.07$\\
\enddata

\tablenotetext{a}{We measured the flux with an aperture size of $1''.2$ and $3''.0$ for \nii 122\um and \nii 205\um, respectively.}
\end{deluxetable*}

From the Band 8 observations, the [NII]~122~$\mu$m line is detected in both of the SMG and the QSO.
The spectra for individual galaxies are shown in Figure~\ref{fig:niiall}.
The line intensity maps are shown in Figure~\ref{fig:nii122} with the peak positions of the \cii line that are consistent with each other.

As listed in Table~\ref{tab:flux}, the width of the \nii122\um line for the QSO is $613 \pm 133$ km s$^{-1}$ which is broader roughly by a factor of two than those observed in the \nii~205\,$\mu$m line ($297\pm 104$  km s$^{-1}$), \cii  ($300\pm28$  km s$^{-1}$) and CO lines ($\approx300-350$  km s$^{-1}$) in the literature (e.g., \citealt{Salome2012, Carniani2013}).
We performed the following tests to verify whether the high line width originates from the systematic errors of the analyses.
First, we did not find systematic differences in the line profile between the tapered and the natural-weight map.
Second, we found that the line profile is robust regardless of continuum-subtraction methods: the continuum subtraction based on the 1-D spectrum using the $0''.6$-aperture is consistent with the $\mathtt{imcontsub}$ and $\mathtt{uvcontsub}$.
Therefore, we conclude that the different line widths between \nii122 and \nii205 for the QSO are likely real.
This may indicate higher electron densities at higher velocities for the QSO that we will discuss in the following section.
\section{Discussion}\label{sec:discussion}
\begin{figure*}
\center
\includegraphics[width = 15 cm, bb = 0 0 1920 800]{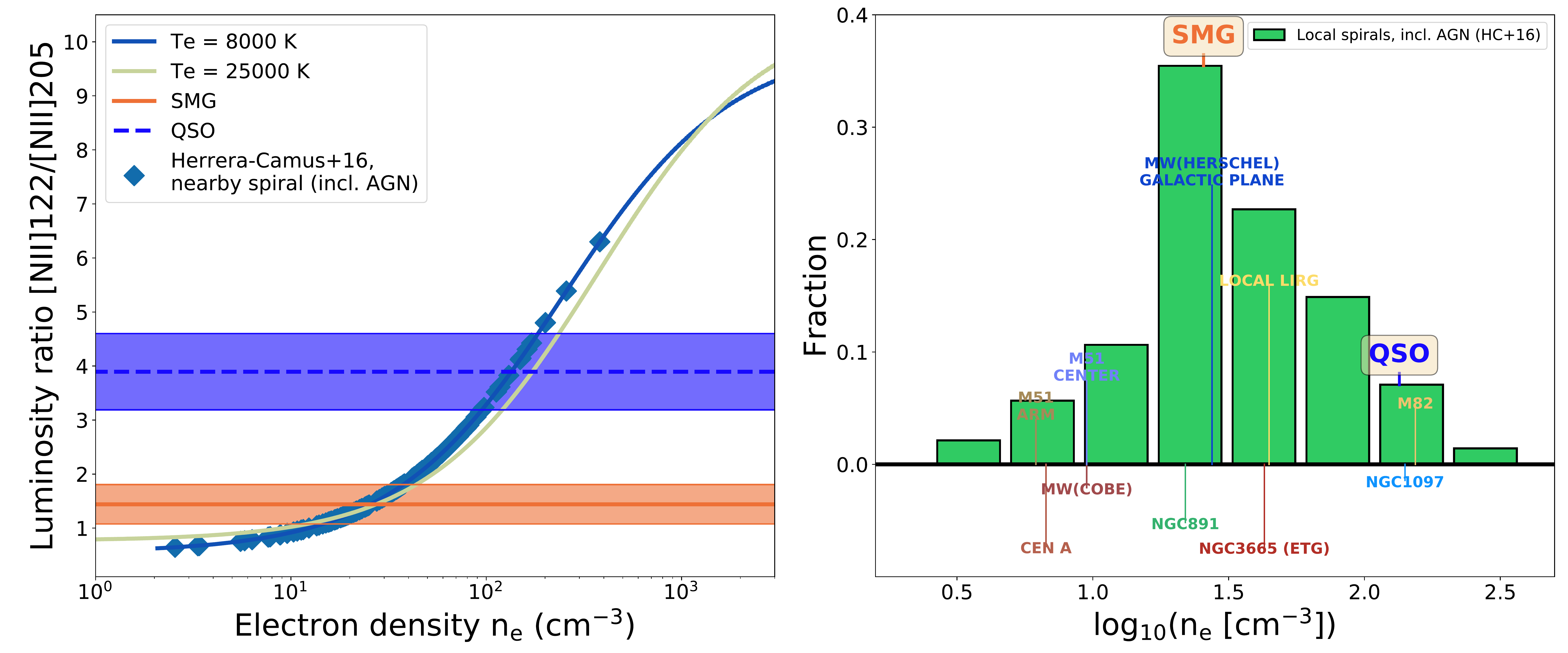}
\caption{Left : The [NII] line luminosity ratio as a function of electron density. The two solid curves are to indicate different assumptions of electron temperatures ($T_{\rm e} = 8000$ K and $25000$ K). The line ratios for the SMG and the QSO are shown as orange solid and blue dashed lines, respectively. The line ratios obtained from local spirals (\citealt{Herrera-Camus2016}) are also plotted in diamonds. Right : The histogram for the distribution of electron density, based on the observed line ratio for \citet{Herrera-Camus2016} to compare with the BR1202-0725 system. The remaining data sets are retrieved from \citealt{Bennett1994} (MW:COBE), \citealt{Goldsmith2015} (MW : galactic plane), \citealt{Parkin2013} (M51), \citealt{Petuchowski1994} (M82), \citealt{Parkin2014} (Cen A), \citealt{Hughes2015} (NGC 891), \citealt{Beirao2012} (NGC1097), \citealt{Xiao2018} (NGC3665, ETG), \citealt{DiazSantos2017} (local LIRGs). \label{fig:niiratio}}
\end{figure*}

We estimate the electron density using the observed ratios between two fine-structure lines of N$^{+}$.
We used the PYNEB package (\citealt{pyneb}) to perform the calculations.
The observed 122\um/205\um line luminosity ratios are $1.44\pm0.36$ and $3.89\pm0.71$ for the SMG and the QSO, respectively.
These correspond to the electron densities of $26^{+12}_{-11}$ (SMG) and $134^{+50}_{-39}$ cm$^{-3}$ (QSO) at the electron temperature of $T_e$ = 8000 K (Figure~\ref{fig:niiratio}), which is used in local spiral galaxy studies (\citealt{Herrera-Camus2016}).

We evaluate potential systematic errors originating from the flux extraction methods in the following manner. First, using the 80k$\lambda$-tapering map, the \nii122\um fluxes are $1.19\pm0.49$ and $1.93\pm0.58$, for the SMG and the QSO, respectively, with the same aperture size of $3''.0$. 
This is consistent with the value obtained from the 330k$\lambda$-tapered maps measurement within errors. 
Second, we also measured the line fluxes using a smaller aperture size of $1''.8$ for the \nii205\um data which is determined after taking into account the emitting size of the \nii122\um line at most ($\approx1''.2$ from the growth curve) and the \nii205\um beam size.  
The flux values from the aperture photometry are $1.00\pm0.07$ and $0.57\pm0.04$ and Jy km s$^{-1}$ for the SMG and the QSO, respectively, providing $n_e = 41^{+17}_{-15}$ (SMG) and $199^{+88}_{-63}$ cm$^{-3}$ (QSO). 
While the estimates above give the lower limit of the electron densities when the size difference between two \nii lines are large (i.e., the \nii122\um-emitting regions being much smaller than the \nii205\um emission), the estimate here from the smaller aperture size serves as a gauge for the central regions. 
Third, if we perform a 2D-gaussian fit for the the \nii122\um emission using the $3''$ aperture with the 330k$\lambda$ map, the flux values are $2.18\pm1.01$ and $1.68\pm0.77$ for the SMG and the QSO, where the uncertainties then become quite large. 
Based on these potential systematic errors, we conclude that the derived electron densities can increase up to by a factor of $\sim1.5$ for the QSO and $\sim3$ for the SMG.

Variation in electron densities in different galaxies have been argued in several studies. 
For example, \cite{Herrera-Camus2016} argued that the electron density correlates with the star formation surface rate density for local spirals using the same \nii tracer, based on the spatially resolved emissions at $\sim$kpc scale.
\cite{Kaasinen2017}, using the optical tracer of \oii for $z\sim1.5$ star forming galaxies, discussed that the SFR is the main driver of varying electron densities.
On the other hand, \citet{Sanders2016} did not find a clear trend of electron density with SFR.
We estimated the rest-frame 123-\um dust continuum (Band 8) sizes based on the $\mathtt{uvmultifit}$ (\citealt{Marti-Vidal2014}), which are $\approx 1$ kpc for both galaxies. 
Considering the similar star-formation rates (SFRs) of $\approx 1000~M_{\odot}$yr$^{-1}$ (e.g., \citealt{Salome2012}) and the similar dust sizes, we do not find the dependence of electron density on the SFR nor SFR surface density on the global scale.

The difference may be indicative of different phases of the black hole growth and/or different gas distributions in the SMG and the QSO.
One possible scenario is that the gas may be more centrally concentrated in the QSO compared to the SMG.
This is counter-intuitive from the preferred formation scenario of elliptical galaxies and the connection between SMGs and QSOs (e.g., \citealt{Hopkins2008b, Toft2014}) where (SMG-like) heavily dust-obscured compact phase with ``denser" ISM precedes the optically bright QSO phase. But, so far no conclusive argument has been made for the connection.
The different \nii line widths in the QSO might indicate higher line ratios of \nii122/205 in the high velocity components, and thus higher electron densities.
If the above scenario is considered, this may be ascribed to gas in the core perhaps at the inner peak of the rotation curve, possibly close to the black hole.
We investigated whether the line profiles are different in the center ($r<0''.2$) and outer region ($0''.2<r<0''.4$). 
But, we could not confirm any statistically significant difference in the fitted line widths partially owing to the low S/N.

Alternatively, the high density gas in the QSO may be a signature of (moderately dense) ionized outflowing gas.
We note that there is a ``red wing" in the \cii line profile (\citealt{Carilli2013b, Carniani2013}), which may be associated with a faint companion or with an outflow.
Observations of AGN-driven galactic outflows in the local universe (e.g., \citealt{Sakamoto2009, Kawaguchi2018}) support the idea of denser gas in the outflowing wind, perhaps due to gas compression.
Since it is difficult to obtain the matched resolution spectra for both galaxies owing to the sensitivity limit, future deeper high resolution observations are needed to confirm.

For more comparison, we compiled the available data sets for various types of galaxies including our Galaxy (MW) and local galaxies as shown in Figure~\ref{fig:niiratio}.
We note that these local measurements are, in most cases, based on spatially resolved emissions and they have a range of electron densities within the galaxies, while our case is for the global average value of the system, assuming that both [NII] lines are coming from the same region. 
The SMG has a comparable electron density compared to those observed in the Galactic Plane (\citealt{Goldsmith2015}) and the average values of nearby, star-forming galaxies (\citealt{Herrera-Camus2016}) using the same tracers, even though the SFRs different by two-to-three orders of magnitude.
Meanwhile, the QSO shows a value comparable to the starburst galaxy like M82 (\citealt{Petuchowski1994}) and NGC 1097 (\citealt{Beirao2012}). 
The value is also similar to typical $n_e$ values found in the central regions of nearby galaxies (\citealt{Herrera-Camus2016}), which are represented by the last two bins in the electron density distribution in the right panel of Figure~\ref{fig:niiratio}.

There are limited number of higher redshift ($z>1$) galaxies with the [NII] line detections for comparison (e.g., \citealt{Zhang2018, Novak2019}).
In \cite{Zhang2018}, they estimated the lower limits of electron densities for lensed, dusty starbursts at a range of $z=1-3.6$ based on stacking analysis, which is $n_e >100$ cm$^{-3}$.
Given the range of electron densities in the BR1202-0725 system, the stacking analysis may have missed a portion of dusty star-forming galaxies with low electron densities like BR1202-0725 SMG.
\cite{Novak2019} also reported a lower limit of electron density ($n_e>180$ cm$^{-3}$) for a QSO at z=7.5, which is higher than our estimate.
Similarly, but using the rest-frame optical lines of \oii and \sii, several studies reported higher electron densities on average compared to local galaxies ranging between $\sim 100-250$ cm$^{-3}$ (e.g., \citealt{Sanders2016, Kaasinen2017}), but there are scatters in the measurements.
It may be worth noting that these optical lines can trace slightly denser gas in the $100 < n_e/{\rm cm}^{-3} < 10^4$ range compared to the \nii lines $10 < n_e/{\rm cm}^{-3} < 500$. Thus, it is likely that rest-frame optical lines $n_e$ measurements (e.g., [OII] and [SII] lines) yield, on average, higher electron density measurements.

In this respect, there may be extremely high electron densities in central regions or elsewhere perhaps with extreme SFR densities where the \nii lines are not viable for measuring electron densities. 
As seen from the resolved measurement in the local galaxies (e.g. \citealt{Goldsmith2015, Herrera-Camus2016}), we do not expect these galaxies to have uniform electron gas densities across the galaxy, but rather to follow some sort of distribution (e.g., a log-normal distribution like the diffuse warm ionized medium). 
If such is the case, the [NII]122/205 line ratio can only probe part of the whole density distribution.
Therefore, it could be that both QSO and SMG have similar high {\it mean} electron densities but different distribution widths, which could be the reason why the [NII]-based $n_e$ measurement in the SMG are lower than that in the QSO. 
To confirm the existence of extremely high density regime, we need other lines instead to trace such regime with higher critical density, such as [NIII] or the combination of [O~{\sc iii}]52$\mu$m and [O~{\sc iii}]88$\mu$m, which can be only accessible from space telescopes. 

Finally, considering the existence of heavily obscured galaxies such as Arp 220 and the fact that shorter wavelengths tend to be more affected by the dust, the reduction of the intrinsic value of the [NII]122/205 line ratio of the SMG compared to QSO might be at least partially, owing to the extremely dusty nature of the SMG. 
Deeper high angular resolution observations at various wavelength would confirm the true nature of the SMG and the QSO.
The success of the \nii 122\um and 205\um detections at $z=4.69$ demonstrate the power of future systematic surveys of extreme starbursts at $z>4$ using these lines for probing the ISM conditions and the effects on surrounding environments in terms of electron densities.

\acknowledgments
We deeply appreciate the anonymous referees for the fruitful discussions and suggestions for the revision of the Letter. We thank Rodrigo Herrera-Camus for providing the data set of local galaxy survey for [NII] lines and fruitful discussions. We also thank Zhi-yu Zhang for the helpful discussions on the treatment of the flux measurement.
SC acknowledges support from the ERC Advanced Grant INTERSTELLAR H2020/740120.
This paper makes use of the following ALMA data: ADS/JAO.ALMA \#2013.1.00259.S. ALMA is a partnership of ESO (representing its member states), NSF (USA) and NINS (Japan), together with NRC (Canada) and NSC and ASIAA (Taiwan) and KASI (Republic of Korea), in cooperation with the Republic of Chile. The Joint ALMA Observatory is operated by ESO, AUI/NRAO and NAOJ.
SC is supported from the ERC Advanced Grant INTERSTELLAR H2020/740120.
This work was supported by NAOJ ALMA Scientific Research Grant
Numbers 2018-09B. R.M. acknowledges ERC Advanced Grant 695671 ``QUENCH”.

\vspace{5mm}
\facilities{ALMA}


\software{astropy \citep{Astropy} 
          }









\end{document}